\documentclass[aps,preprint,prl, superscriptaddress, showpacs,nobibnotes]{revtex4-2}
\usepackage{graphicx}
\usepackage{bm,amsmath}
\usepackage{dcolumn}
\usepackage{natbib,xcolor,float}
\usepackage{mhchem,bm}
\usepackage{placeins}
\usepackage{ragged2e}
\usepackage{setspace}
\makeatletter
\def\@fnsymbol#1{\ensuremath{\ifcase#1\or \dagger\or \dagger,*\or
   \mathsection\or \mathparagraph\or \|\or **\or \dagger\dagger
   \or \ddagger\ddagger \else\@ctrerr\fi}}
   \makeatother
\newcommand{\upcite}[1]{\textsuperscript{\textsuperscript{\cite{#1}}}}

\begin{document}
\title{Magnetic flat band in metallic kagome lattice FeSn}
\author{Yaofeng Xie}
\affiliation{Department of Physics and Astronomy, Rice University, Houston, Texas, USA}
\author{Lebing Chen}
\affiliation{Department of Physics and Astronomy, Rice University, Houston, Texas, USA}
\author{Tong Chen}
\affiliation{Department of Physics and Astronomy, Rice University, Houston, Texas, USA}

\author{Qi Wang}
\affiliation{Department of Physics and Beijing Key Laboratory of Opto-Electronic Functional Materials \& Micro-Nano Devices, Renmin University of China, Beijing, China}
\author{Qiangwei Yin}
\affiliation{Department of Physics and Beijing Key Laboratory of Opto-Electronic Functional Materials \& Micro-Nano Devices, Renmin University of China, Beijing, China}
\author{J. Ross Stewart}
\affiliation{ISIS Facility, STFC Rutherford-Appleton Laboratory, Didcot, UK}
\author{Matthew B. Stone}
\affiliation{Neutron Scattering Division, Oak Ridge National Laboratory, Oak Ridge, TN 37831, USA}
\author{Luke L. Daemen}
\affiliation{Neutron Scattering Division, Oak Ridge National Laboratory, Oak Ridge, TN 37831, USA}
\author{Erxi Feng}
\affiliation{Neutron Scattering Division, Oak Ridge National Laboratory, Oak Ridge, TN 37831, USA}
\author{Huibo Cao}
\affiliation{Neutron Scattering Division, Oak Ridge National Laboratory, Oak Ridge, TN 37831, USA}
\author{Hechang Lei}
\affiliation{Department of Physics and Beijing Key Laboratory of Opto-Electronic Functional Materials \& Micro-Nano Devices, Renmin University of China, Beijing, China}
\author{Zhiping Yin}
\affiliation{Center for Advanced Quantum Studies and Department of Physics, Beijing Normal University, Beijing, China}
\author{Allan H. MacDonald}
\affiliation{Department of Physics, University of Texas-Austin, Austin, Texas, USA}
\author{Pengcheng Dai}
\affiliation{Department of Physics and Astronomy, Rice University, Houston, Texas, USA}
\date{\today}

\begin{abstract}
\textbf{
In two-dimensional (2D) metallic kagome lattice materials, destructive interference of electronic hopping pathways around the kagome bracket can produce nearly localized electrons, and thus electronic bands that are flat in momentum space.  When ferromagnetic order breaks the degeneracy of the electronic bands and splits them into the spin-up majority and spin-down minority electronic bands, quasiparticle excitations between the spin-up and spin-down flat bands should form a narrow localized spin-excitation Stoner continuum coexisting with well-defined spin waves in the long wavelengths.  Here we report inelastic neutron scattering studies of spin excitations in 2D metallic Kagome lattice antiferromagnetic FeSn and paramagnetic CoSn, where angle resolved photoemission spectroscopy experiments found spin-polarized and nonpolarized flat bands, respectively, below the Fermi level.  Although our initial measurements on FeSn indeed reveal well-defined spin waves extending well above 140 meV coexisting with a flat excitation at 170 meV, subsequent experiments on CoSn indicate that the flat mode actually arises mostly from hydrocarbon scattering of the CYTOP-M commonly used to glue the samples to aluminum holder. Therefore, our results established the evolution of spin excitations in FeSn and CoSn, and identified an anomalous flat mode that has been overlooked by the neutron scattering community for the past 20 years.
}
\end{abstract}

\maketitle

Magnetic metals support both collective spin-wave [Fig. 1(a)] and quasiparticle (Stoner) [Figs. 1(b,c)] spin-flip excitations\upcite{heisen,lovesey,stoner1,stoner2,slater,wohlf,fawc}.   In three-dimensional (3D) metallic ferromagnets like iron\upcite{lynn,perring} and nickel\upcite{mook},  the quasiparticle bands are broad and spin waves are well-defined only at long wavelengths, disappearing when they enter the Stoner continuum at intermediate spin-wave momenta [Fig. 1(c)]\upcite{kirs1,kirs2,moriya}.  In 2D metallic kagome lattice materials [Figs. 1(d-g)], destructive interference of electronic hopping pathways around the kagome bracket can produce nearly localized electrons, and thus electronic bands that are flat in momentum space [Figs. 1(h-j)]\upcite{suther,leykam,tang,neupert,ghim}.  When combined with spin-orbit coupling and magnetic order, the geometric frustration induced flat bands provide an ideal platform for novel topological phases\upcite{guo,mazin,chen}, ferromagnetism\upcite{tasaki}, and superconductivity\upcite{bist,cao}.  Although flat bands below the Fermi level have recently been identified in the antiferromagnetic (AF) kagome metal FeSn\upcite{kang1}, paramagnetic kagome metal CoSn\upcite{liu,kang2}, as well as in other materials\upcite{ye,lin,yin1}, their influence on spin-wave and Stoner excitations is  unknown.  

In general, spin-flip excitations in a magnet can be interpreted in terms of either a quantum spin models\upcite{heisen,lovesey} with local moments on each atomic site [Fig. 1(a)], or a Stoner\upcite{stoner1,stoner2,slater,wohlf,moriya} itinerant electron model.  In insulating ferromagnets such as EuO, magnetic excitations can be fully described by a Heisenberg Hamiltonian\upcite{dietrich} with spins on Eu lattice sites.   In ferromagnetic metals, magnetic order breaks the degeneracy of the electronic bands, splitting spin-up majority and spin-down minority electrons [Fig. 1(b)]\upcite{wohlf}.  For 3D metallic FM Fe and Ni, the low-energy spin waves are strongly damped when they enter a broad Stoner continuum of band-electron spin-flips that extends over several eV in energy [Fig. 1(c)]\upcite{lynn,perring,mook,kirs1,kirs2}.  For a paramagnetic metal, there is no splitting of the degenerate electronic bands, and one would not expect to observe a Stoner continuum\upcite{moriya}.  In strongly correlated materials like copper and iron-based superconductors, the subtle balance between electron kinetic energy and short-range interactions can lead to debates concerning whether magnetism has a localized or itinerant origin\upcite{keimer,dai1}. 

In some 2D crystals, electrons can be confined in real space to form flat bands, for example through geometric lattice frustration\upcite{guo,mazin,chen}.  The flat bands of magic-angle twisted bilayer graphene\upcite{bist} provide one example of this route toward strong electronic correlation\upcite{cao}.  The kagome lattice depicted in Fig. 1(h), in which the simplest nearest neighbor electronic hopping model predicts destructive phase interference [Fig. 1(i)] leading to real-space electron localization and a flat electronic band [Fig. 1(j)]\upcite{mazin}, provides a second.  Recently, a spin-polarized flat electronic band has been identified in the AF kagome metal FeSn at an energy E = $230\  \mathrm {\pm}\ 50$ meV below the Fermi level by angle-resolved photoemission spectroscopy (ARPES) experiments\upcite{kang1}.  FeSn is a A-type AF with antiferromagnetically coupled FM planes\upcite{sales}, which we will view as 2D ferromagnets.  Neutrons should in principle detect the electron-hole-pair Stoner excitations from the majority-spin flat band below the Fermi level to minority-spin bands near or above the Fermi level [Fig. 1(b)]\upcite{wohlf,moriya}.  Since neutron scattering measures electron-hole-pair excitations, having a flat spin-up electronic band below the Fermi level is a necessary, but not a sufficient condition to observe a flat Stoner continuum band.  Instead, such a dispersionless narrow energy spin excitation band also requires a flat spin-down electronic band above (or near) the Fermi level\upcite{moriya}. Unfortunately, ARPES measurements cannot provide any information concerning  such an electronic band above the Fermi level, although density functional theory (DFT) calculations suggest its presence\upcite{kang1}.  For comparison, although ARPES measurements have also identified flat band at an energy E = $270\  \mathrm {\pm}\ 50$ meV below the Fermi level in CoSn\upcite{liu,kang2}, one would not expect to observe a flat Stoner continuum band due to degenerate electronic bands and paramagnetic nature of the system\upcite{meier}.

In this paper, we report inelastic neutron scattering (INS) studies of spin excitations in 2D metallic Kagome lattice antiferromagnetic FeSn\upcite{sales} and paramagnetic CoSn\upcite{meier}.  For FeSn, our initial measurements reveal well-defined spin waves extending well above 140 meV and a narrow ~24 meV wide band of excitations that cannot be described by a simple spin-wave model.  While these data suggest the presence of electron-hole-pair Stoner excitations from the majority-spin flat band below the Fermi level to minority-spin bands near or above the Fermi level in FeSn, subsequent experiments on paramagnetic CoSn also have the same flat mode coexisting with expected paramagnetic spin excitations. Through careful analysis of INS spectra under different conditions, we conclude that the observed flat mode actually arises mostly from hydrocarbon scattering of the CYTOP-M commonly used to glue the samples to aluminum holder\upcite{rule}. Therefore, our results established the evolution of spin excitations in FeSn and CoSn, and identified an anomalous flat mode that has been overlooked by the neutron scattering community for the past 20 years.

\section{Results} 
We have carried out INS experiments to study spin waves and search for anomalous Stoner excitations in AF kagome metallic FeSn\upcite{sales} and paramagnetic CoSn\upcite{meier}.  The structure of FeSn consists of 2D kagome nets of Fe separated by layers of Sn, and exhibits AF order below $T_N \approx $ 365 K with in-plane FM moments in each layer stacked antiferromagnetically along the c-axis [Fig. 1(d)]\upcite{sales}.  Since each unit cell contains three Fe atoms [Fig. 1(e)], we expect one acoustic and two optical spin-wave branches in a local moment Heisenberg Hamiltonian\upcite{sales,xing,chisnell}.  Figures 1(f,g) show the reciprocal spaces corresponding to the crystal structures of FeSn depicted in Figs. 1(d,e), respectively. CoSn has the same crystal structure as that of FeSn but is paramagnetic at all temperatures\upcite{meier}.

Figure 2(a) illustrates low-energy spin excitations observed in FeSn along the high symmetry directions in the ($H$,$K$) plane depicted in Fig. 1(g).  We see a nearly isotropic spin-wave mode stemming from the $\Gamma$ point at the FM zone center that moves towards the zone boundary with increasing energy.  Along the [$H$, 0, 0] direction towards the M point, the acoustic mode reaches the zone boundary at around 90 meV.  The left panel of Figure 2(b) shows an image of the wave vector transfer (\textbf{Q}) and energy (E) dependence of spin excitations around the M point.  The data confirm that the acoustic spin waves reach around 90 meV at the M point, and reveal an optical spin-wave mode that is visible at energies between 115 and 140 meV, and is characterized by out-of-phase oscillations amongst the three Fe sublattice spins.  The evolution of the in-plane \textbf{Q}-dependence of acoustic spin waves with increasing energy is summarized in Fig. 2(d), revealing spin wave rings around the $\Gamma$ point that enlarge with increasing energy.  Unlike the in-plane FM spin waves stemming from the $\Gamma$ point, which have excitation energies exceeding 140 meV [Figs. 2(a,d)],  the spin waves along the c-axis stem from the AF ordering wave vectors $\rm{\boldsymbol{Q}}$ = (0, 0, 0.5+$L$) with ($L$ = 0, 1) and reach the zone boundary around E $\approx$ 20 meV [Fig. 2(f)], reflecting relatively weak coupling between ferromagnetic kagome planes.  We also observe an easy-axis anisotropy gap $\mathit{\Delta}_a \approx$ 2 meV due to single-ion magnetic anisotropy [Fig. 2(h)]\upcite{sales}. 

To understand these observations, we start with a local moment Heisenberg Hamiltonian $H=\sum_{i\neq j}{{J_{ij}\boldsymbol{S}}_i{\cdot \boldsymbol{S}}_j}+\ \sum_i{A{({{\boldsymbol{S}}_i}^x)}^2}$, where $J_{ij}$ is magnetic exchange coupling of the spin ${\boldsymbol{S}}_i$ and ${\boldsymbol{S}}_j$,  the exchange coupling \textit{J} consists of \textit{J${}_{1}$} (in-plane nearest neighbor), \textit{ J${}_{2}$} (in-plane next-nearest neighbor), and \textit{J${}_{c}$} (out-of-plane nearest neighbor) [Figs. 1(d,e)]\upcite{sales}.  The magnetic anisotropy is represented by the second term parameterized by \textit{A}.  Since spin waves have FM in-plane dispersion [Figs. 2(a,d)] and AF out-of-plane dispersion [Figs. 2(f,h)], we assume \textit{J${}_{1}$},\textit{ J${}_{2}$ }$\mathrm{<}$ 0 and \textit{J${}_{c}$ }$\mathrm{>}$ 0.  Assuming an in-plane spin easy-axis along the [3.72,1,0] (\textit{x}) direction\upcite{yamaguchi,kulsh}, we define A ($\mathrm{<}$ 0) to be the single-ion anisotropy.  Figures 2(c,e,g) shows the outcome of our least-square fit to the spin-wave excitations of FeSn in Figs. 2(a,d,f), respectively.  The solid lines in the right panel of Fig. 2(b) and in Fig. 2(f) are spin wave dispersions determined from the Heisenberg Hamiltonian with  \textit{J${}_{1}$} = -20.7$\ \mathrm {\pm}$ 3.5 meV, \textit{J${}_{2}$} = -5.1$\ \mathrm{\pm}$ 1.3 meV, \textit{J${}_{c}$} =  9.5$\ \mathrm {\pm}$ 0.2 meV, \textit{A} = -0.052$\ \mathrm {\pm}$ 0.002 meV,, and Fe spin $S=1$.  The experimental in-plane FM exchange couplings obtained from this fit are smaller than theoretical predictions, while the the c-axis exchange coupling is larger by a factor of two\upcite{sales}.

Figures 3(a,b) show 2D \textbf{Q}-\textit{E} space images measured along the in-plane [\textit{H},0] and [\textit{H},\textit{H}] directions over a broader energy regime, and reveal a clear \textbf{Q}-independent excitation continuum centered at \textit{E} =  173$\ \mathrm{\pm}$ 1 meV with an energy width of about 24 meV.  Energy cuts at different wave vectors within the first Brillouin zone reveal similar results, indicating that the mode is indeed \textbf{Q}-independent [Fig. 3(c)].  Fig. 3(c) reveals that the intensity of the mode decreases with increasing \textbf{Q}.  Figures 3(d,e) show the 2D \textbf{Q}-\textit{E} space images along the in-plane [\textit{H},0] and [\textit{H},\textit{H}] directions, calculated using the Heisenberg Hamiltonian magnetic exchange parameters from the fit in Fig. 2.  Although the energy of the expected second optical spin waves is similar to that of the flat mode, the experimental in-plane \textbf{Q}-dependence in Figs. 3(a,b,c,g) differs qualitatively from the theoretical  Heisenberg Hamiltonian fits in Figs. 3(d,e,f).  For example, the optical spin waves from a Heisenberg Hamiltonian should have no magnetic scattering intensity within the first Brillouin zone [Fig. 3(f)], in clear contrast with the experimental data [Fig. 3(g)].  

To determine whether or not the magnetic excitations of FeSn can be understood within a Heisenberg Hamiltonian with $S=1$\upcite{sales}, we consider the energy dependence of the local dynamic susceptibility ${\chi }^{''}(E)$  [Fig. 3(h)], obtained by integrating the imaginary part of the generalized dynamic spin susceptibility ${\chi }^{''}(\boldsymbol{Q},E)$ over the first Brillouin zone [the green shaded region in Fig. 1(g)] at different energies\upcite{dai2} using ${\chi }^{''}\mathrm{(}\boldsymbol{\mathrm{Q}},E\mathrm{)\ =\ }\left[\mathrm{1-exp}\left(\mathrm{-}\frac{E}{k_BT}\right)\right]S(\boldsymbol{Q},E)$, where $S(\boldsymbol{Q},E)$ is the measured magnetic scattering in absolute units, $E$ is the neutron energy transfer, and $k_B$ is the Boltzmann’s constant.  Since the static ordered moment per Fe is $M\approx 1.85\ {\mu }_B$ at 100 K\upcite{sales}, the total magnetic moment $M_0$ of FeSn satisfies  $M^2_0=M^2+\left\langle m^2\right\rangle \approx 5.5\ {\mu }^2_B$ per Fe, where the local fluctuating moment $\left\langle m^2\right\rangle \approx 2\ {\mu }^2_B$ is obtained by integrating ${\chi }^{''}(E)$ at energies below 150 meV.  In the local moment Heisenberg Hamiltonian with S=1, the total moment sum rule implies that $M^2_0=g^2S(S+1)$, where $g\approx 2$  is the Lande g-factor, requiring a fluctuating moment contribution of $g^2S=4$ ${\mu }^2_B$ per Fe, which is a factor of 2 larger than the measured $\left\langle m^2\right\rangle \approx 2\ {\mu }^2_B$ per Fe.  The solid and dashed lines in Fig. 3(h) are the calculated ${\chi }^{''}(E)$ in absolute units assuming $S=1$, and 0.5, respectively.  These unusual fluctuation properties suggest that electronic itineracy contributes plays a role in both the observed static ordered moment of FeSn\upcite{sales}.  We remark that flat spin wave flat bands can occur in kagome lattice ferromagnets with Dzyaloshinskii-Moriya (DM) interactions\upcite{xing}, but these have an entirely different origin\upcite{chisnell}. 

If the flat band observed in Figs. 3(a,b) is indeed magnetic and arises from a flat spin-down electronic band above (or near) the Fermi level [Fig. 1(b)], there should not be a flat band in paramagnetic CoSn due to the absence of spin polarized flat electronic bands. Surprisingly, excitation spectra in CoSn again show a flat mode at $\sim$ 170 meV [Fig. 4(a)], suggesting that the mode may not have a magnetic origin.  Since our FeSn and CoSn samples are glued on the aluminum plates by CYTOP-M which is an amorphous fluoropolymer but contains one hydrogen to facilitate bonding to metal surface\upcite{rule,cytop}, the hydrogenated amorphous carbon films formed between sample and aluminum plates should have C-H bending and stretching vibrational modes occurring around 150-180 meV and 350-380 meV, respectively\upcite{heitz,john}.  Indeed, a scan with incident neutron beam energy of 500 mV reveals the presence of a flat mode around 350-380 meV [Fig. 4(b)], thus confirming the nonmagnetic nature of the flat modes at 170 meV and 360 meV.

To compare the paramagnetic scattering in CoSn and spin waves in FeSn, we first determine the energy scales of the phonon scattering.  The blue and red data points in Figs. 4(c) and 4(d) show energy dependence of the scattering at $\rm{Q} = 7.7-8.3 \rm{\AA}^{-1}$ and $ 1.7-2.3 \rm{\AA}^{-1}$, respectively.  At large Q, the scattering is dominated by phonons and shows a cut off above 50 meV for both samples.  Figures 4(e) and 4(f) compare wave vector dependence of the scattering above the phonon cut off, which reveal broad paramagnetic scattering centered around the $\Gamma$ point in CoSn and clear spin waves in FeSn.  These results suggest that the energy scale of the paramagnetic scattering in CoSn is comparable to spin waves of FeSn.       

Although our measurements in CoSn conclusively identified the flat modes at 170 meV arises from C-H bending vibrational modes, these results cannot establish conclusively that the mode arises from CYTOP-M and there is still the possibility that the 170 meV flat mode in FeSn (Fig. 3) has a magnetic component.  To test these possibilities, we deposited liquid CYTOP-M on aluminum plates but without sample, and baked the assembly in a vacuum furnace at temperatures similar to how FeSn samples were mounted with CYTOP-M.  The experimental outcome at 5 K reveals a clear flat mode at 170 meV, thus establishing that solidified CYTOP-M gives rise to the scattering [Figs. 5(a) and 5(b)].  The presence of a flat mode at 360 meV in Fig. 5(c) confirms that the scattering at 170 meV arises from the C-H bending mode\upcite{heitz,john}.  Figure 5(d) shows the scattering from liquid CYTOP-M at room temperature, where the 170 meV C-H bending mode in solidified CYTOP-M moves to 150 meV.

To further test if pure FeSn without CYTOP-M can also be contaminated by hydrocarbons, we prepared fresh single crystals of FeSn and carried out measurements at 5 K using unaligned single crystals on SEQUOIA.  We find weak and broad excitations at 170 meV and 360 meV, respectively, suggesting the presence of additional hydrocarbon contamination.   By shifting the incident beam neutron away from the sample using a motorized mask, the hydrocarbon contamination is still present with the similar intensity ratio between 170 meV and 360 meV modes (Table 1).  Our careful infrared absorption spectrum analysis on the thermal shielding suggests the presence of hydrocarbon contamination, probably a silicone oil accidentally contaminating the vacuum system at SEQUOIA.  Therefore, we conclude that the observed scattering at 170 meV in FeSn arises mostly from solid CYTOP-M with small additional contamination from hydrocarbons on thermal shielding.  

\section{Discussion}

To account for electronic itineracy, we calculate the electronic structure of FeSn in the paramagnetic and AF ordered states using a combination of DFT and dynamical mean field theory (DFT+DMFT)\upcite{kotliar}.  In the paramagnetic state, the mass enhancements of the Fe 3\textit{d} electrons near the Fermi level are about 1.8 for the $d_{z^2}\mathrm{\ }$orbital, 2.4 for $d_{x^2-y^2}$ and $d_{xy}$ orbitals, and around 4 for the $d_{xz}$ and $d_{yz}$ orbitals [Fig. 4(a)].  Since these values are similar to the values in iron arsenide superconductors,  we conclude that FeSn is a Hund's metal\upcite{yin2} with intermediate strength correlations. Figure 4(b) shows the DFT+DMFT electronic structures, with the dominant Fe \textit{d} orbitals highlighted, in the paramagnetic state.  To calculate the spin polarized magnetically ordered state, we assume FM exchange couplings along both the in-plane and \textit{c}-axis directions because of the small \textit{c}-axis AF exchange coupling [Fig. 2(f)] and the simplicity of the electronic band structure in a ferromagnet. We find that the mass enhancements of the Fe 3\textit{d} electrons near the Fermi level are reduced substantially  to around 1.5 for all Fe 3\textit{d} orbitals in the minority spin channel, about 1.7 for the $d_{z^2}$ and $d_{xy}$ orbitals, and 2.4 for the $d_{x^2-y^2}$, $d_{xz}$ and $d_{yz}$ orbitals in the majority spin channel.  Figures 4(c,d) respectively show the calculated spin-up and spin-down electronic structure of FeSn in the FM ordered state.  In addition to confirming the spin-up flat electronic band around 230 meV below the Fermi level seen in ARPES experiments\upcite{kang1}, we obtain a flat minority spin electronic band around 170 meV above the Fermi level.  

Based on ARPES measurement and DMFT calculation, we would expect the presence of a flat band Stoner continuum arising from quasiparticle excitations between the spin-up majority and spin-down minority electrons.  Our failure to observe such a mode means that it is much weaker than spin waves in FeSn.  From our absolute intensity measurements of spin waves in FeSn, we find that a local moment Heisenberg model with $S = 1$ cannot account for the integrated spin wave intensity, suggesting the important role played by itinerant electrons in these materials.     

Although CYTOP-M has been used by neutron scattering community as a glue to mount small samples for over 20 years, its characteristics at high energies have not been reported\upcite{rule}.  This is mostly because the difficulty in carrying INS at energies above 100 meV at traditional reactor sources.  The development of neutron time-of-flight measurements at spallation sources allows measurements at energies well above 200 meV, and the flat mode was missed in previous work\upcite{dai2} because of its weak intensity and its weak Q dependence.  Our identification of the flat C-H bending and stretching vibrational modes should help future neutron scatterers to separate these scattering from genuine magnetic signal.   

\section{Methods}
\subsection{{\bf Sample synthesis, structural and composition characterization.}}

Single crystals of FeSn and CoSn were grown by the self-flux method. The high-purity Fe (Co) and Sn were put into corundum crucibles and sealed into quartz tubes with a ratio of Fe (Co) : Sn = 2 : 98.  The tube was heated to 1273 K and held there for 12 h, then cooled to 823 K (873 K) at a rate of 3 (2) K/h.  The flux was removed by centrifugation, and shiny crystals with typical size about 2$\mathrm{\times}$2$\mathrm{\times}$5 mm${}^{3}$ can be obtained.  The single crystal X-ray diffraction (XRD) pattern was performed using a Bruker D8 X-ray diffractometer with Cu \textit{K${}_{\alpha }$ }radiation (\textit{$\lambda$} = 0.15418 nm) at room temperature (Fig. S1). 

The elemental analysis was performed using energy-dispersive X-ray (EDX) spectroscopy analysis in a FEI Nano 450 scanning electron microscope (SEM).  In order to determine composition of FeSn accurately, we carefully polished FeSn surface using sandpaper and carried out EDX measurements on five FeSn crystals (Fig. S2).  The average stoichiometry of each crystal was determined by examination of multiple points (5 positions).  As shown in Table S1, the atomic ratio of Fe:Sn is close to 1:1.

To further determine the crystalline quality and stoichiometry of the samples used in neutron scattering experiments, we took X-ray single-crystal diffraction experiments on two pieces of these samples at the Rigaku XtaLAB PRO diffractometer housed at Spallation Neutron Source at Oak Ridge National Laboratory (ORNL).  The measured crystals were carefully suspended in Paratone oil and mounted on a plastic loop attached to a copper pin/goniometer (Fig. S3).  The single-crystal X-ray diffraction data were collected with molybdenum K radiation (\textit{$\lambda$}\ =\ 0.71073\ {\AA}).  More than 2800 diffraction Bragg peaks were collected and refined using Rietveld analysis (Table S2).  We find no evidence of superlattice peaks indicating possible Fe vacancy order (Fig. S3). The refinement results indicate less than 1.5\% possible Fe vacancy (Fig. S4), suggesting that the single crystals are essentially fully stoichiometric.

To determine whether the AF phase transition in our sample is consistent with earlier work\upcite{sales}, we carried out temperature and field dependence of the magnetization measurement.  Figure S5 reveals an AF phase transition around 377 K under a magnetic field of 0.5 and 14 T [Figs. S5(a) and (b)].  Magnetic field dependences of the magnetization for both field directions at various temperatures [Fig. S5(c) and (d)] are also in agreement with earlier work\upcite{sales}.

\subsection{{\bf Neutron scattering experiments.}}

INS measurements on FeSn were carried out using the MAPS time-of-flight chopper spectrometer at the ISIS Spallation Neutron Source, the Rutherford Appleton Laboratory, UK\upcite{ewings}. INS measurements on CoSn and FeSn are also performed using the SEQUOIA spectrometer at the Spallation Neutron Source, Oak Ridge National Laboratory\upcite{gran}. Fifty pieces of single crystals of FeSn with the total mass of 0.97 g were co-aligned on one single piece of aluminum plate and mounted inside a He displex.  Figure S6 shows that the mosaic of aligned single crystals is about 6 degrees.  The crystal structure of FeSn is hexagonal with space group P${}_{6}$/mmm with lattice parameters $a=b=5.529\ \textrm{\AA}$, and $c=4.4481\ \textrm{\AA}$\upcite{sales}. The lattice parameters of CoSn are $a=b=5.528\ \textrm{\AA}$, and $c=4.26\ \textrm{\AA}$\upcite{meier}. We define the momentum transfer \textbf{Q} in 3D reciprocal space in ${\textrm{\AA}}^{-1}$ as $\boldsymbol{Q}=H{\boldsymbol{a}}^*+K{\boldsymbol{b}}^*+L{\boldsymbol{c}}^*$, where \textit{H}, \textit{K}, and \textit{L} are Miller indices and ${\boldsymbol{a}}^{\boldsymbol{*}}=2\pi (\boldsymbol{b}\times \boldsymbol{c})/[\boldsymbol{a}\cdot \left(\boldsymbol{b}\times \boldsymbol{c}\right)]$, ${\boldsymbol{b}}^{\boldsymbol{*}}=2\pi (\boldsymbol{c}\times \boldsymbol{a})/[\boldsymbol{a}\cdot \left(\boldsymbol{b}\times \boldsymbol{c}\right)]$, ${\boldsymbol{c}}^{\boldsymbol{*}}=2\pi (\boldsymbol{a}\times \boldsymbol{b})/[\boldsymbol{a}\cdot \left(\boldsymbol{b}\times \boldsymbol{c}\right)]$ with $\boldsymbol{a}=a\widehat{\boldsymbol{x}}$, $\boldsymbol{b}={a(\mathrm{cos} 120\ }\widehat{\boldsymbol{x}}+{\mathrm{sin} 120\ \widehat{\boldsymbol{y}}),\ }$ $\boldsymbol{c}=c\widehat{\boldsymbol{z}}$ [Figs. 1(d,e)].  The horizontal scattering plane is [\textit{H},0,\textit{L}] and INS data were collected with incident neutron energies set to $E_i=300,\ 250,\ 150,\ 100,\ 30,$ and 15 meV in Horace mode and the temperature is set at 5 K.  The neutron scattering data is normalized to absolute units using a vanadium standard, which has an accuracy of approximately 30\%\upcite{dai2}.  

\subsection{{\bf Heisenberg model fitting to spin waves of FeSn.}}

We use the Heisenberg model and least-square method to fit spin waves of FeSn (Figs. S7-S9).  The software package used was SpinW and Horace\upcite{toth}.  The Heisenberg Hamiltonian is $H=\sum_{i\neq j}{{J_{ij}\boldsymbol{S}}_i{\cdot \boldsymbol{S}}_j}+\ \sum_i{A{({{\boldsymbol{S}}_i}^x)}^2}$ as discussed in the main text.  Note in our Heisenberg Hamiltonian fit to spin wave data, we only used dispersion relations from experiments, and assumed $S=1$, which is close to the 1.86 ${\mu }_B$ per Fe ordered moment\upcite{sales}.   The overall intensity from SpinW fit, when considered in absolute units, is considerably higher than the experiment [Fig. 2(h)].  This suggests that Heisenberg Hamiltonian overestimates the spin wave intensity contribution from the ordered moment.  We first determine the interlayer coupling \textit{J${}_{c}$}.  Using linear spin wave theory, we find that the spin-wave band top along the \textit{c} axis direction to have energy $E_{Ltop}=2|J_c|$ when anisotropy \textit{A} is not too large.  A constant-\textbf{Q} cut at the zone boundary wave vector $\boldsymbol{Q}=(0,0,1.75)$ shows \textit{E${}_{L}$${}_{top}$} = 19.0 $\mathrm{\pm}$ 0.3 meV [Fig. 2(f)], therefore yielding \textit{J${}_{c}$} = 9.5 $\mathrm{\pm}$ 0.2 meV.  

To estimate \textit{A}, we note that the spin gap of ${\mathit{\Delta}}_a\approx 2$ meV at the AF wave vector $\boldsymbol{Q}=(0,0,0.5)$ [Fig. 2(h)].  Several types of anisotropy are considered as shown in Fig. S7, and we choose the dipole-like anisotropy because it is the only one that can create a gap in the neutron spectrum.  Assuming \textit{J${}_{c}$} = 9.5 meV, we find ${\mathit{\Delta}}_a=8.761\sqrt{\left|A\right|}$. The calculated anisotropic parameter is\textit{ A }= 0.052 $\mathrm{\pm}$ 0.002 meV.  The presence of \textit{A} will cause the \textit{E${}_{Ltop}$ }to shift by 0.1 meV, and therefore can be safely ignored in fits of the in-plane magnetic exchange couplings.

To fit the in-plane exchanges \textit{J${}_{1}$} and\textit{ J${}_{2}$}, we use constant-\textit{E} cuts along the [\textit{H},0] and   [\textit{H},\textit{H}] direction made for spin wave energies of 15-80 meV, Gaussian fits are used to extract the spin-wave peak position.  We iterate over [\textit{J${}_{1}$},\textit{ J${}_{2}$}], calculate every dispersion, and calculate the square error between calculation and experiment of each [\textit{J${}_{1}$},\textit{ J${}_{2}$}] configuration (Fig. S7).  To narrow down the range of \textit{J${}_{1}$} and\textit{ J${}_{2}$}, we fit part of the optical spin wave modes at the \textit{M} point (Fig. S8).  A constant-\textbf{Q} cut integrating between [\textit{H}, \textit{H}] = (-0.1, 0.1), [\textit{H}, 0] = [0.4, 0.6], and $L$ = [-1, 3] is used as fitting data [Figs. 2(c,e,g)].  Least-square error mode is used to get the best fit, giving \textit{J${}_{1}$} = -20.7$\ \mathrm{\pm}\ $3.5meV and \textit{J${}_{2}$} = -5.1$\ \mathrm{\pm}\ $1.3meV, which concludes the Heisenberg model fitting of the data. 

Note that the error bars of these parameters are estimated as follows: First, calculate the least square error using best-fit parameter $J_0$ and denote it as $R_0$; then determine the parameter $J^{\prime}_{0}$ when the square error values give 2$R_0$, and the error bar is given by $\Delta J = | J^{\prime}_0 - J_0 |$.  This error range gives 68\% confidence interval, meaning that if the residual of the fit has a Gaussian distribution, the calculation generated by range $[J_0-\Delta J, J_0+\Delta J]$ can cover 68\% (i.e. 1$\sigma$ in Gaussian distribution) of the data points.

\subsection{{\bf DFT+DMFT calculations.}}

The electronic structures and spin dynamics of FeSn in the paramagnetic and magnetically ordered states are computed using DFT+DMFT method\upcite{kotliar}.  The density functional theory part is based on the full-potential linear augmented plane wave method implemented in Wien2K\upcite{blaha}.  The Perdew--Burke--Ernzerhof generalized gradient approximation is used for the exchange correlation functional\upcite{perdew}.  DFT+DMFT was implemented on top of Wien2K and was described in detail before\upcite{haule1}.  In the DFT+DMFT calculations, the electronic charge was computed self-consistently on DFT+DMFT density matrix.  The quantum impurity problem was solved by the continuous time quantum Monte Carlo (CTQMC) method\upcite{haule2,werner} with a Hubbard \textit{U} = 4.0 eV and Hund's rule coupling \textit{J} = 0.7 eV in both the paramagnetic state and the magnetically ordered state.  Bethe-Salpeter equation is used to compute the dynamic spin susceptibility where the bare susceptibility is computed using the converged DFT+DMFT Green's function while the two-particle vertex is directly sampled using CTQMC method after achieving full self-consistency of DFT+DMFT density matrix\upcite{yin3}.  For the magnetically ordered state, the averaged Green's function of the spin up and spin down channels is used to compute the bare susceptibility. In the paramagnetic state, an electronic flat band of dominating $d_{xz}$ and $d_{yz}$ orbital characters locates a few meV above the Fermi level (Fig.4b).  In the magnetic state, the spin exchange interaction leads to about 400 meV splitting of this electronic flat band where the spin-up flat band is pushed down to about -230 meV (Fig.4c) whereas the spin-down flat band is pushed up to $\mathrm{\sim}$170 meV (Fig.4d). The experimental crystal structure (space group P6/\textit{mmm}, No. 191) of FeSn with lattice constants $a=b=5.297\ \textrm{\AA}$ and$\ c=4.481\ \textrm{\AA}$ is used in the calculations. Figure S10 shows orbital-resolved band structures of FeSn in the paramagnetic, spin up, and spin down magnetically ordered state.  We also note that the possible $\mathrm{\sim}$1.2\% iron deficiency in FeSn obtained from X-ray refinement (Table S2) is not expected to modify the band structure. 

\subsection{\bf The infrared absorption measurements.} To determine if the thermal heat shielding of SEQUOIA acquired an organic coating, we cut a small piece of the shielding right after the experiment and carried out the infrared absorption spectrum measurement on that piece.  The spectrum in Fig. S11 shows clear evidence of an organic coating on the foil exposed to the contaminant. The modes at 1072, 1038, and 960 $\rm{cm}^{-1}$ in the fingerprint region of the spectrum point unambiguously at Si-O-Si stretches (siloxane groups). The strong peak at 1128 $\rm{cm}^{-1}$ can also be assigned to Si--R (R = aryl) stretches.  The very strong, sharp mode at 1240 $\rm{cm}^{-1}$ is characteristic of a symmetric deformation mode of CH3 in Si-CH3 bonds. The expected frequency range is 1240-1290 $\rm{cm}^{-1}$. Electropositive metals attached to Si tend to move this peak to higher frequencies, whereas for siloxanes, the band occurs at the lower end. The mode at 1462 $\rm{cm}^{-1}$ is either a scissoring -CH2 mode or, more likely, the asymmetric deformation mode of -CH3. The presence of C-H bonds is confirmed by the existence of a strong absorption band with three clear, relatively sharp modes just below 3000 $\rm{cm}^{-1}$ (symmetric and antisymmetric C-H stretches). This information points at the contaminant being a polysiloxane material, probably a silicone oil accidentally contaminating the vacuum system at SEQUOIA. The origin of this contaminant is uncertain: lubricant or electrical insulating material around power or signal cables. The very intense mode at 1728 $\rm{cm}^{-1}$ is undoubtedly a carbonyl stretching mode. It may be part of a pendant group on the polysiloxane backbone (e.g., a carbinol group used to modify the hydrophobic/hydrophilic of the oil). No further attempt was made at solving the precise molecular structure of the polymeric material.

\subsection{{\bf Data availability.}}The data that support the plots within this paper and other findings of this study\textbf{ }are available from the corresponding authors upon reasonable request.

\subsection{{\bf Code availability.}}The codes used for the DFT+DMFT calculations in this study are available from the corresponding authors upon reasonable request.

\section{Acknowledgement}
First and foremost, we wish to express our sincere appreciation to the anomalous referees who reviewed this paper, particularly referee 2.  In the original draft of the paper, we only have data for FeSn. It is the comment of referee 2 that inspired us to carry out measurements on CoSn, resulting in the discovery of the C-H bending and stretching vibrational modes from solid CYTOP-M unknown to the community for the past 20 years. The referee 2 also prevented us from making a major mistaken in identifying the C-H bending mode as a flat Stoner continuum. Although the outcome is disappointing, it nevertheless reveals that having a spin polarized flat electronic band does not necessitate the observation of a magnetic flat band.  The neutron scattering work at Rice was supported by US NSF-DMR-1700081 and by the Robert A. Welch Foundation under Grant No. C-1839 (P.D.).  Z.P.Y. was supported by the NSFC (Grant No. 11674030), the Fundamental Research Funds for the Central Universities (Grant No. 310421113), the National Key Research and Development Program of China grant 2016YFA0302300.  The calculations used high performance computing clusters at BNU in Zhuhai and the National Supercomputer Center in Guangzhou.  H.L. was supported by the National Key R\&D Program of China (Grants No. 2018YFE0202600, 2016YFA0300504), the National Natural Science Foundation of China (No. 11774423, 11822412), the Beijing Natural Science Foundation (Grant No. Z200005), the Fundamental Research Funds for the Central Universities, and the Research Funds of Renmin University of China (RUC) (18XNLG14, 19XNLG17).  A.H.M. was supported by the U.S. Department of Energy, Office of Science, Basic Energy Sciences, under Award \# DE‐SC0019481, and by the Welch Foundation under award TBF1473. E.F. and H.C. acknowledge the support of U.S. DOE BES Early Career Award KC0402010 under contract DE-AC05-00OR22725.  A portion of this research used resources at the Spallation Neutron Source, a DOE Office of Science User Facility operated by the Oak Ridge National Laboratory.

\section{Author contributions}

P.D. and H.L. conceived the project.  Q.W., Q.Y., and H.L. prepared the samples.  Neutron scattering experiments were carried out by Y.F.X and J.R.S., and analyzed by Y.F.X., L.C., with help from T.C. and P.D.  Single crystal X-ray diffraction measurements were carried out by E.F and H.C. DFT+DMFT calculations were carried out by Z.P.Y.  The paper was written by P.D., Y.F.X, Z.P.Y., L.C., H.L. and A.H.M..  All authors made comments.  The authors declare no competing financial interests. 
Correspondence and requests for materials should be addressed to P.D. (pdai@rice.edu) or Z.P.Y. (yinzhiping@bnu.edu.cn) or H.L. (hlei@ruc.edu.cn) or A.H.M. (macdpc@physics.utexas.edu).

\newpage

\begin{figure}[h]
	\includegraphics[width = 0.5\columnwidth]{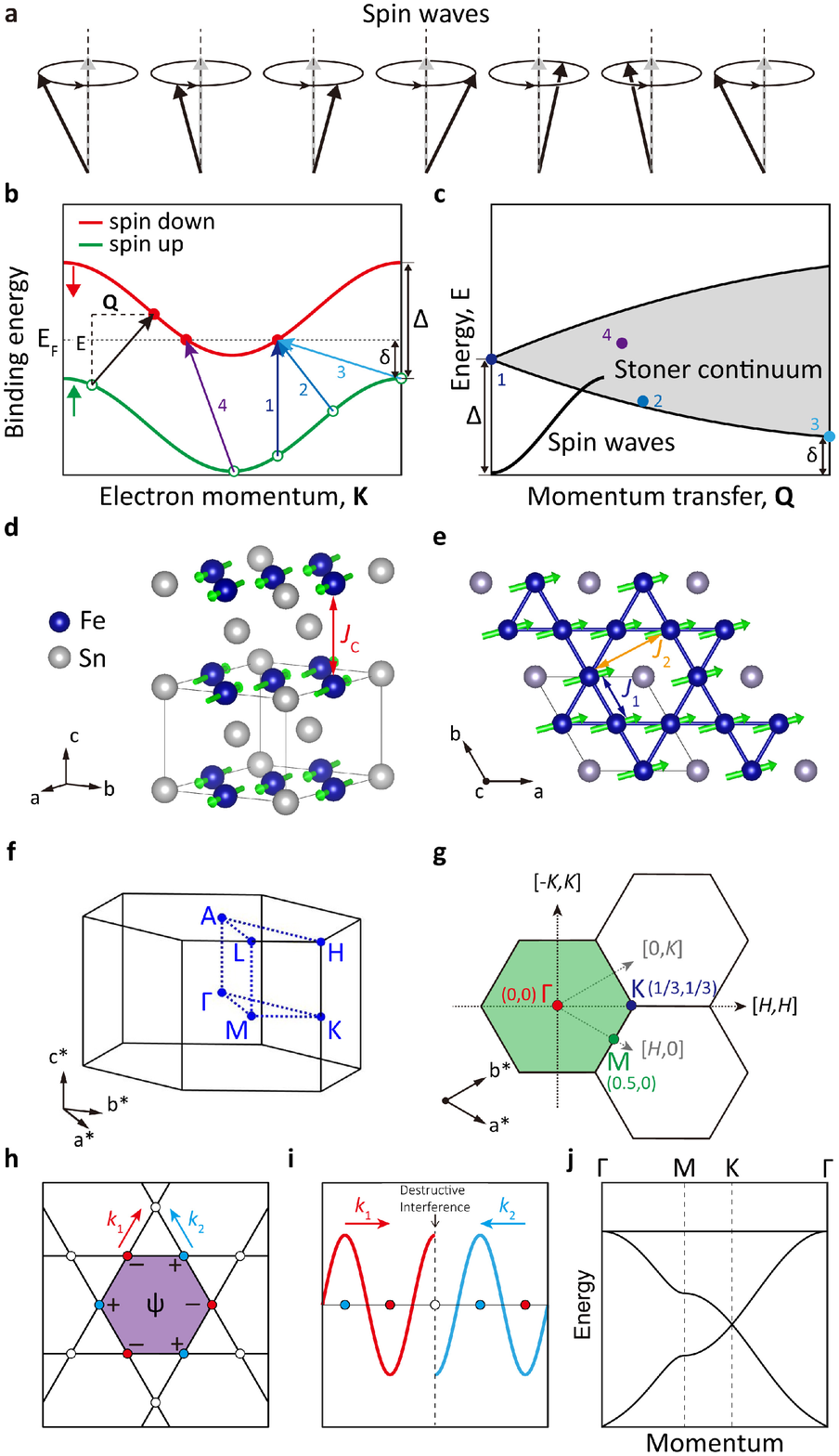}
	\singlespace
	\caption{\label{fig-1} \textbf{Schematic diagrams of spin waves, Stoner continuum, lattice and magnetic structures of FeSn and its reciprocal space notations.} (a) Collective spin waves in a local moment Heisenberg model.  (b,c) Schematic illustrations of Stoner excitations. This excitation corresponds to a transition from occupied states in a spin-up band to unoccupied states in a spin-down band.  The numbered arrows in (b) indicate the possible scattering processes, corresponding to the numbered points in (c), the energy spectrum of the single particle electron-hole pair excitations. (d) Crystal and magnetic structures of FeSn. (e) Top view of the kagome lattice, where $J_1$, $J_2$, and $J_c$ are the dominant magnetic interactions used in the Heisenberg model to fit spin waves of FeSn. (f,g) 3D and 2D Brillouin zone of FeSn, respectively. The high symmetry points are specified and the green-colored hexagon marks the integration area in reciprocal space to obtain the total magnetic scattering within one Brillouin zone. (h,i,j) Schematic illustrations of destructive quantum interferences induced electronic confinement and flat band in kagome lattice. Electron hoppings outside the purple-colored hexagon will be prohibited due to the antiphase of flat band eigenstates at the neighboring sublattices, resulting in the perfect electron localization and flat electronic band in reciprocal space.}
\end{figure}

\begin{figure}[p]
	\includegraphics[width = 0.8\columnwidth]{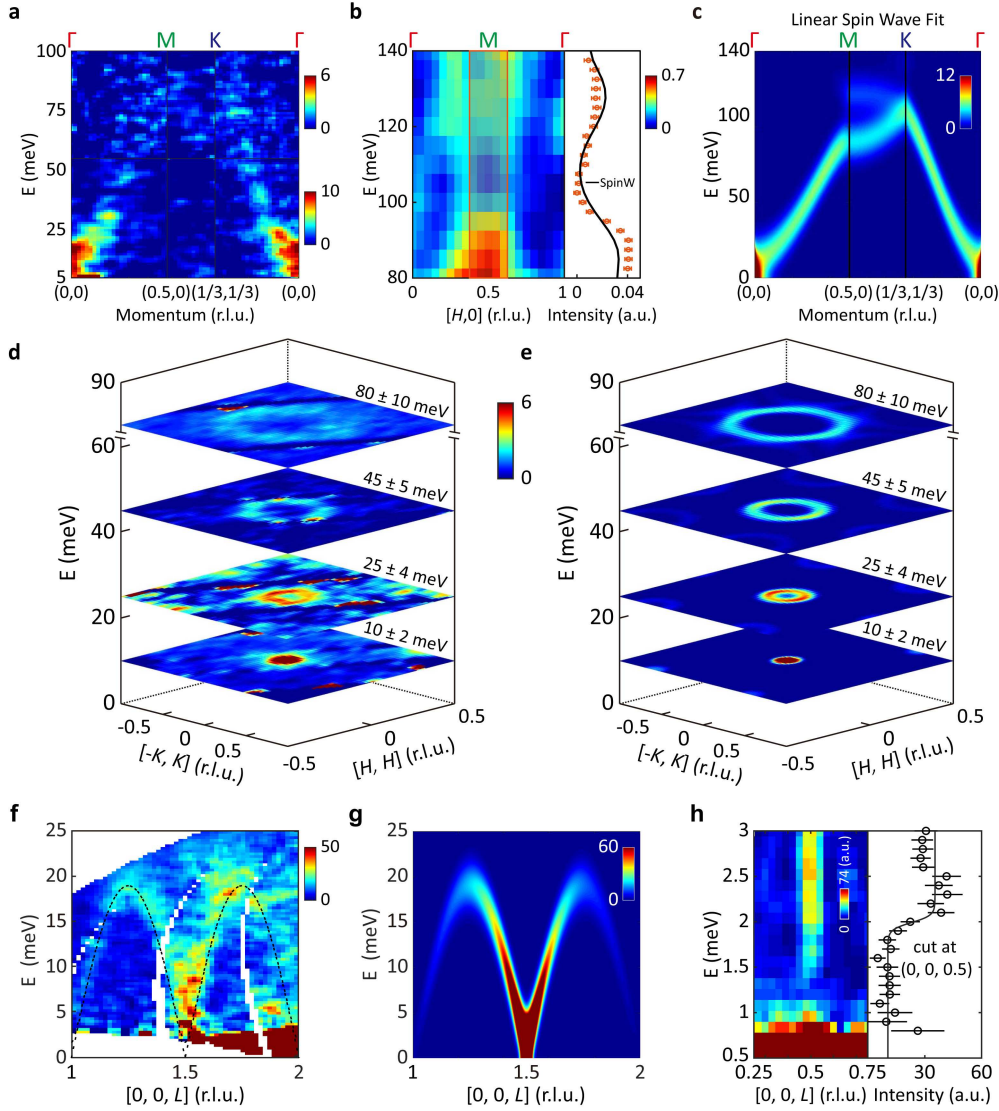}
	\caption{\label{fig-2} \textbf{Low-energy spin waves of FeSn and fit to a Heisenberg Hamiltonian.} (a) Neutron scattering function $S(\boldsymbol{Q},E)$ at 5 K along the high symmetric line \textit{$\mathit{\Gamma}$-M-K-$\mathit{\Gamma}$} through the Brillouin zone. The Data below and above $E=55$ meV were collected with incident neutron energy $E_i=100$  and 250 meV, respectively. (b) Measured $S(\boldsymbol{Q},E)$ at 5 K along the high symmetric line \textit{$\mathit{\Gamma}$-M-$\mathit{\Gamma}$}. The orange box shows the integration range along the [\textit{H},0] direction for the energy cut on the right-hand side.  (c) Calculated $S(\boldsymbol{Q},E)$ along the \textit{$\mathit{\Gamma}$-M-K-$\mathit{\Gamma}$} directions using a Heisenberg model.  (d) Constant-energy cuts within the (\textit{H},\textit{K}) plane at energies $E=10\pm 2,\ 25\pm 4,\ 45\pm 5,\ 80\pm 10$ meV. (e) Constant-energy cuts of the fitted $S(\boldsymbol{Q},E)$ in the Heisenberg model. (f) Measured $S(\boldsymbol{Q},E)$ along the [0,0,\textit{L}] direction. Dashed lines are from the Heisenberg model fit. (g) Calculated $S(\boldsymbol{Q},E)$ along the [0,0,\textit{L}] direction. (h) Images and cuts of magnetic excitations near the AF zone center $\boldsymbol{Q}=(0,0,0.5)$. In the present study, the color bars represent the vanadium standard normalized absolute magnetic excitation intensity in the units of mbarn meV${}^{-1}$ per formula unit, unless otherwise specified.  The calculated spin wave intensity in (c,e,g) is in absolute units assuming $S=1$ in the SpinW+Horace program\upcite{toth}. The error bars in (h) represent statistical errors of 1 standard deviation.}
\end{figure}

\begin{figure}[h]
	\includegraphics[width =0.95\columnwidth]{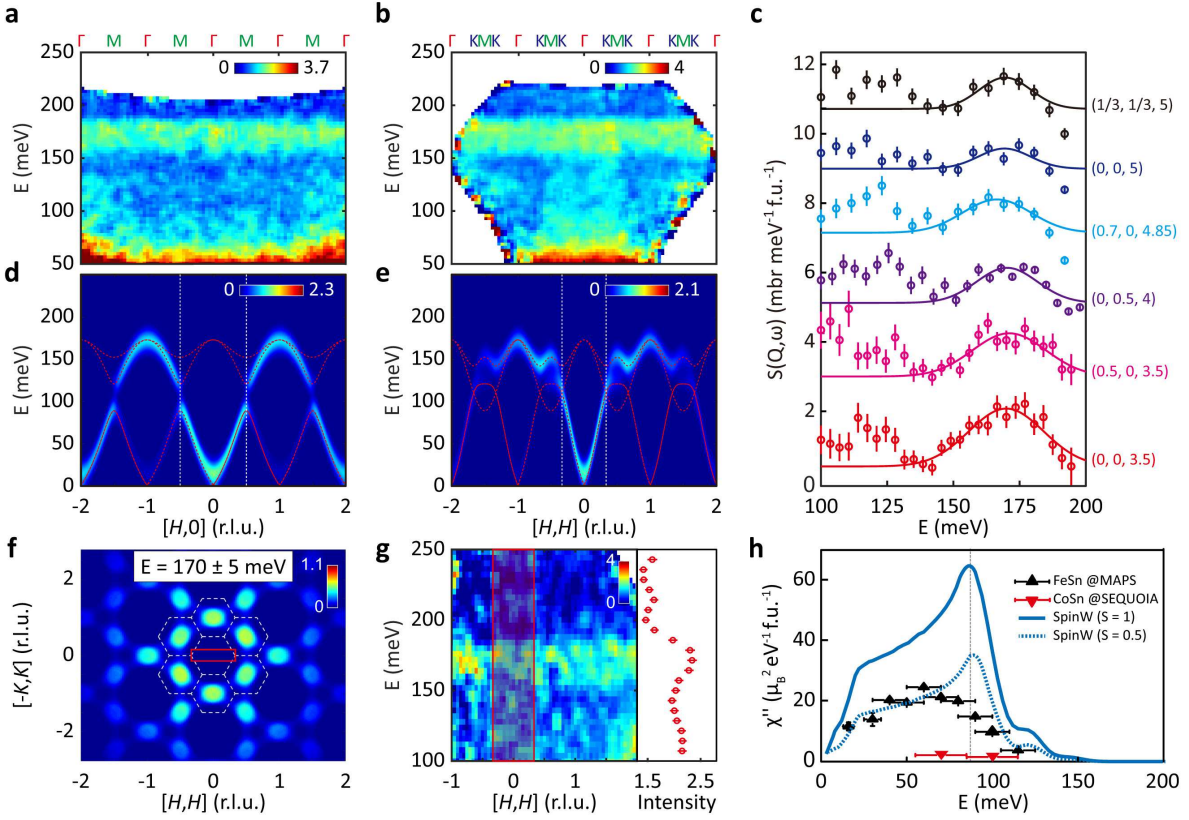}
	\caption{\label{fig-3} \textbf{Flat band like excitations in FeSn.} (a,b) Measured $S(\boldsymbol{Q},E)$ of FeSn at 5 K along the [\textit{H},\textit{H}] and [\textit{H},0] directions, respectively. Data were collected with incident neutron energy of $E_i=250$ meV. The dispersionless magnetic excitation is peaked at around $E=173$ meV with the full width at half maximum (FWHM) of 24 meV. The instrument energy resolution at $E=173$ meV is about 11 meV. (c) Energy cuts at the selected wave vectors within the first Brillouin zone. The scattering seems to decrease in intensity with increasing $\boldsymbol{Q}$. (d,e), Calculated $S(\boldsymbol{Q},E)$ using the Heisenberg model with the same parameters as those in Fig. 2. Red dashed lines are corresponding spin wave dispersions, showing one acoustic and two optical spin waves. Vertical white dashed lines indicate the first Brillouin zone. (f) Constant energy cut of the calculated $S(\boldsymbol{Q},E)$ within the (\textit{H},\textit{K}) plane at $E=170\pm 5$ meV from Heisenberg model. The dashed lines are the Brillouin zone boundaries of the 2D reciprocal lattice. (g) Measured $S(\boldsymbol{Q},E)$ along the [\textit{H},\textit{H}] direction with incident neutron energy of $E_i=300$ meV and corresponding energy cut on the right-hand side, whose $\boldsymbol{Q}$-integration range over \textit{H} and \textit{K} directions is indicated by the red box in (f). (h) Energy-dependent local dynamic susceptibility ${\chi }^{\prime\prime}(E)$ of FeSn, obtained by $\boldsymbol{Q}$-integration of ${\chi }^{\prime\prime}(\boldsymbol{Q},E)$ in the green box of 1(g) and from $1\le L\le 2$. The blue solid and dashed lines are the calculated ${\chi }^{\prime\prime}(E)$ using identical $\boldsymbol{Q}$-integration from the Heisenberg model assuming $S=1$ and $0.5$, respectively.  The statistical errors in (h) do not include the 30\% uncertainty due to the vanadium standard normalization.  The horizontal and vertical error bars in (c) and (h) represent $\boldsymbol{Q}(E)$ integration range and statistical errors of 1 standard deviation, respectively. }
\end{figure}

\begin{figure}[h]
	\includegraphics[width = 0.8 \columnwidth]{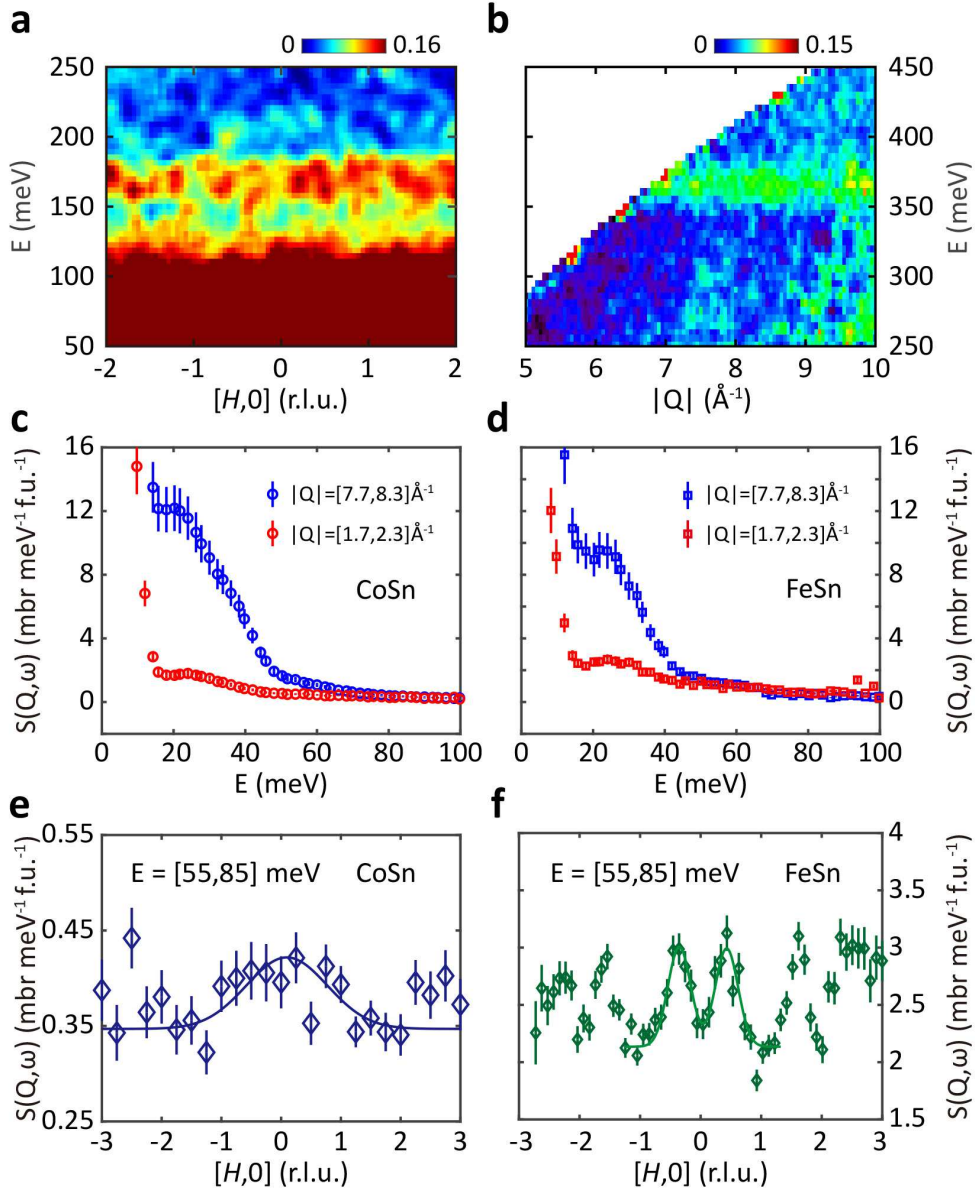}
	\caption{\label{fig-4} \textbf{Measurements of excitations in CoSn.}  (a) Measured $S(\boldsymbol{Q},E)$ of CoSn with $E_\mathrm{i}$ = 300 meV at 5 K along the [$H$,0] direction. (b) Powder-averaged spectrum of CoSn with $E_\mathrm{i}$ = 550 meV at 5 K. (c,d) Comparison between low-momentum-transfer $1.7\le Q \le 2.3\ \mathrm{\AA}^{-1}$ and high-momentum-transfer $7.7\le Q \le 8.3\ \mathrm{\AA}^{-1}$ energy dependence of measured $S(\boldsymbol{Q},E)$ for CoSn (c) and FeSn (d), respectively. (e,f) Momentum dependence of $S(\boldsymbol{Q},E)$ along the [H,0] direction at $E$ = 70 ± 15 meV for CoSn (e) and FeSn (f), respectively. The vertical error bars in (c-f) represent statistical errors of 1 standard deviation.}
\end{figure}

\begin{figure}[h]
	\includegraphics[width = 0.8 \columnwidth]{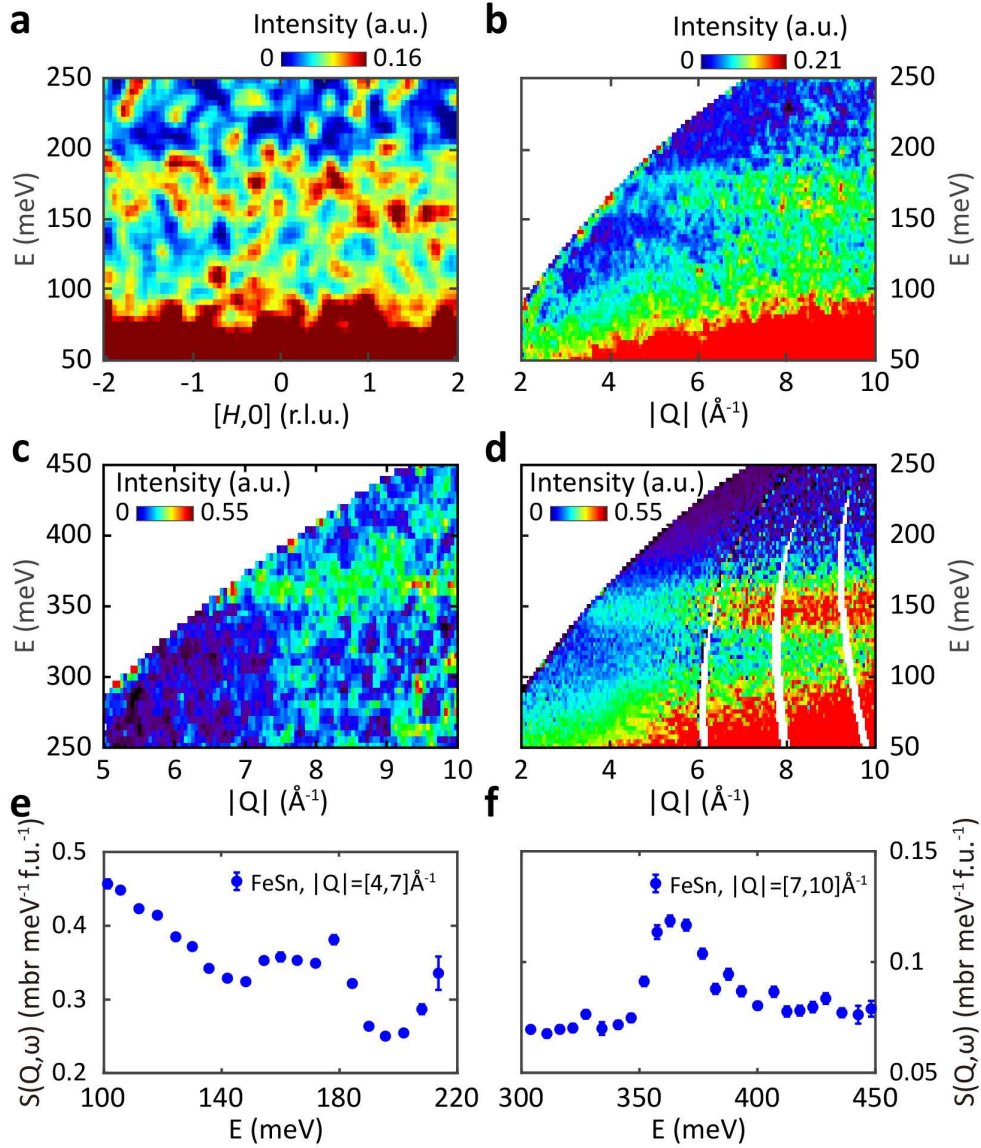}
	\caption{\label{fig-4} \textbf{Measurements of excitations in solidified/liquid CYTOP-M and clean FeSn.} (a,b,c) Inelastic neutron scattering intensity from baked CYTOP-M on aluminum plates at 5 K with $E_\mathrm{i}$ = 300 (a,b) and 550 (c) meV. (d) Inelastic neutron scattering intensity from liquid CYTOP-M at room temperature with $E_\mathrm{i}$ = 300 meV. (e,f) Energy dependence of measured $S(\boldsymbol{Q},E)$ of clean FeSn for momentum transfer $4\le Q \le 7\ \mathrm{\AA}^{-1}$ (e) and $7\le Q \le 10\ \mathrm{\AA}^{-1}$ (f) at 5 K. The vertical error bars in (e,f) represent statistical errors of 1 standard deviation.}
\end{figure}

\begin{figure}[h]
	\includegraphics[width = 0.8 \columnwidth]{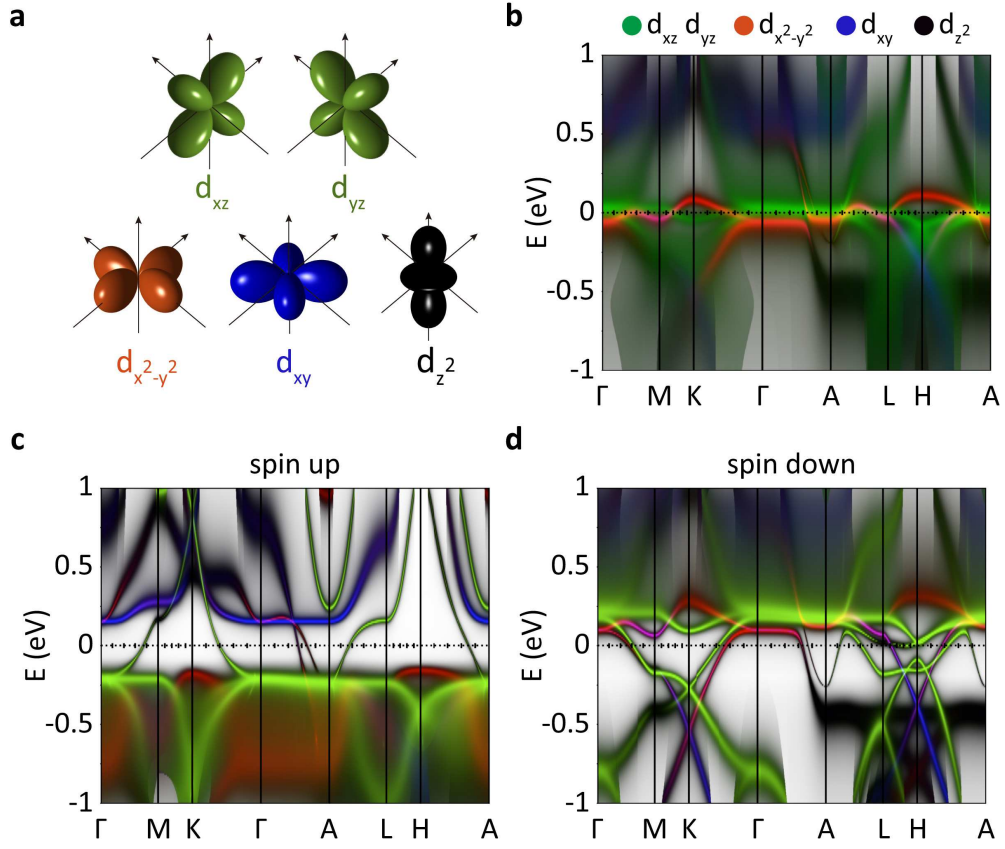}
	\caption{\label{fig-4} \textbf{Electronic structures of FeSn from the DFT + DMFT calculations.} (a) Wave functions of the Fe five \textit{d} orbitals. (b,c,d) Orbital-resolved band structures of FeSn in the paramagnetic, spin up, and spin down magnetically ordered state, respectively. Green color is the contribution from the $d_{xz}$ and $d_{yz}$ orbitals. Red, blue, and black colors represent $d_{x^2-y^2}$, $d_{xy}$, and $d_{z^2}$ orbitals, respectively.}
\end{figure}

\end{document}